\documentclass{article}

\PassOptionsToPackage{numbers, compress}{natbib}
\usepackage[final]{nips_2016}

\usepackage[utf8]{inputenc} % allow utf-8 input
\usepackage[T1]{fontenc}    % use 8-bit T1 fonts
\usepackage{hyperref}       % hyperlinks
\usepackage{url}            % simple URL typesetting
\usepackage{booktabs}       % professional-quality tables
\usepackage{amsfonts}       % blackboard math symbols
\usepackage{nicefrac}       % compact symbols for 1/2, etc.
\usepackage{microtype}      % microtypography

\usepackage{amsmath}
\usepackage[ruled]{algorithm2e} % For algorithms
\usepackage{makecell}
\usepackage{multirow}% http://ctan.org/pkg/multirow
\usepackage{hhline}% http://ctan.org/pkg/hhline
\usepackage{caption}
\usepackage{tabularx}

\usepackage{graphicx} % for EPS, load graphicx instead

\SetAlFnt{\small}
\SetAlCapFnt{\small}
\SetAlCapNameFnt{\small}
\SetAlCapHSkip{0pt}
\IncMargin{-\parindent}

\title{Automatic Identification of Non-Meaningful Body-Movements and What It Reveals About Humans}

\author{M. Iftekhar Tanveer\\
Electrical and Computer Engineering\\
University of Rochester\\
Rochester, NY - 14627\\
USA\\
\texttt{itanveer@cs.rochester.edu}\\
\And
RuJie Zhao\\
Computer Science\\
University of Rochester\\
Rochester, NY - 14627\\
USA\\
\texttt{rzhao2@cs.rochester.edu}\\
\And
Mohammed (Ehsan) Hoque\\
Computer Science\\
University of Rochester\\
Rochester, NY - 14627\\
USA\\
\texttt{mehoque@cs.rochester.edu}\\
}

\begin{document}

\maketitle

\begin{abstract}
We present a framework to identify whether a public speaker's body movements are meaningful or non-meaningful (``\emph{Mannerisms}'') in the context of their speeches. In a dataset of 84 public speaking videos from 28 individuals, we extract 314 unique body movement patterns (e.g. pacing, gesturing, shifting body weights, etc.). Online workers and the speakers themselves annotated the meaningfulness of the patterns. We extracted five types of features from the audio-video recordings: disfluency, prosody, body movements, facial, and lexical. We use linear classifiers to predict the annotations with AUC up to 0.82. Analysis of the classifier weights reveals that it puts larger weights on the \emph{lexical features} while predicting self-annotations. Contrastingly, it puts a larger weight on \emph{prosody features} while predicting audience annotations. This analysis might provide subtle hint that public speakers tend to focus more on the verbal features while evaluating self-performances. The audience, on the other hand, tends to focus more on the non-verbal aspects of the speech. The dataset and code associated with this work has been released for peer review and further analysis.
\end{abstract}

% <concept_id>10003120.10003138.10003140</concept_id>
% <concept_desc>Human-centered computing~Ubiquitous and mobile computing systems and tools</concept_desc>
% <concept_id>10003752.10010070</concept_id>
% <concept_desc>Theory of computation~Theory and algorithms for application domains</concept_desc>

% \ccsdesc[500]{Human-centered computing~Ubiquitous and mobile computing systems and tools}
% \ccsdesc[500]{Theory of computation~Theory and algorithms for application domains}


% We no longer use \terms command
%\terms{Design, Algorithms, Performance}

% \keywords{Behavioral Analysis; Public Speaking; Mannerisms; Data Analysis}

% 

% Author's addresses: M. Tanveer, Electrical and Computer Engineering, University of
% Rochester, itanveer@cs.rochester.edu; R. Zhao and M. Hoque, Computer Science, University of Rochester, \{rzhao2,mehoque\}@cs.rochester.edu}

%\renewcommand{\shortauthors}{M. Tanveer et al.}

%%%%%%%%%%%%%%%%%%%%%%%%%%%%%%% Body of the paper %%%%%%%%%%%%%%%%%%%%%%%%%%%%%%%%%%%%%
\section{Introduction}
\begin{figure}
\centering
 \includegraphics[width=0.5\linewidth]{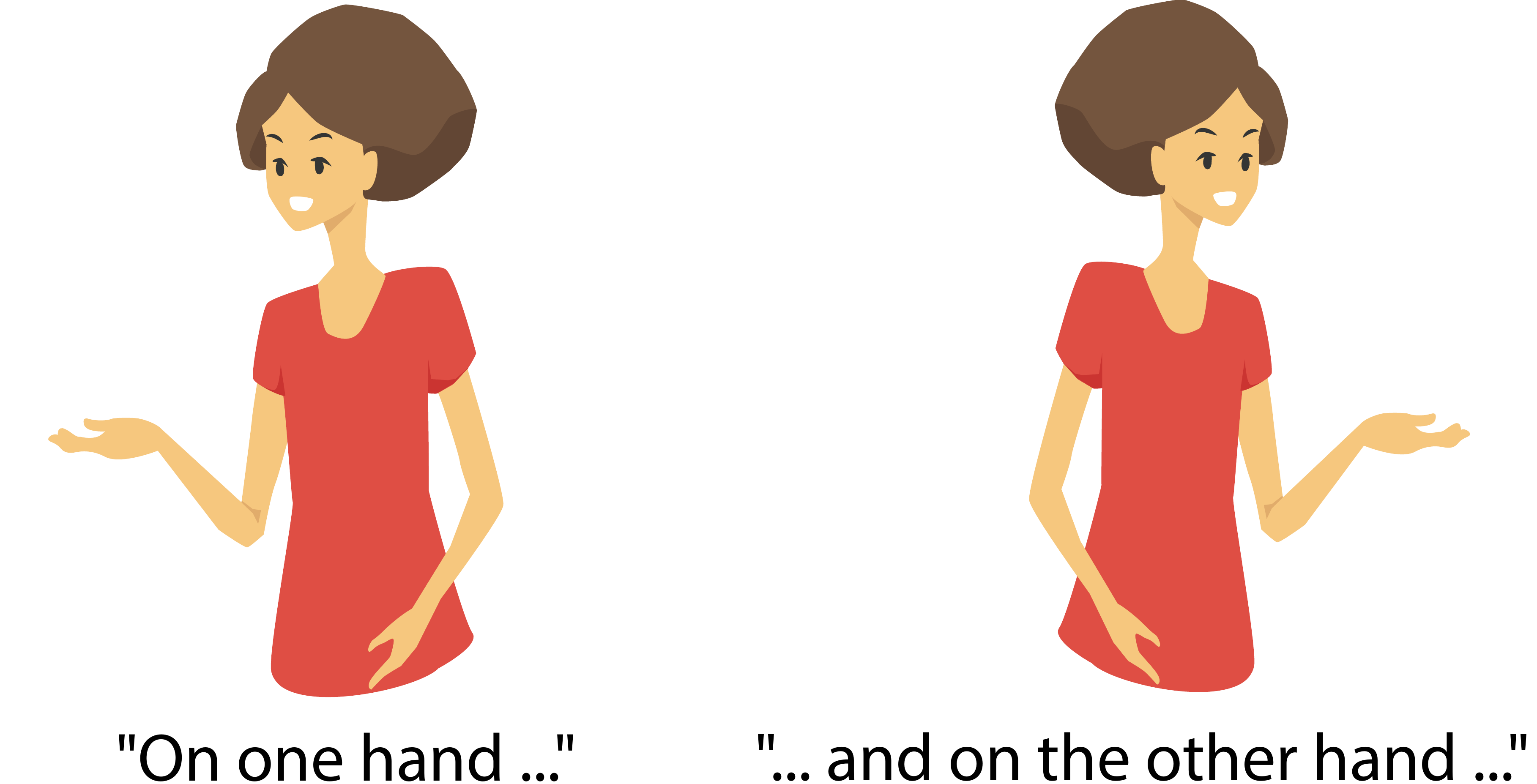}
      \caption{A comparison gesture}
      \label{fig:comparison_gest}
\hfill
\end{figure}

Public speakers use body language to augment verbal content. A hand gesture, for instance, accompanying ``this big'' may convey a sense of size. Gestures also help to convey relational information. For instance, speakers tend to gesture when comparing two things, as illustrated in Figure~\ref{fig:comparison_gest}. These are meaningful body movements---reinforcing the communication by adding to or modifying the content of speech. Accordingly, public speaking experts encourage these nonverbal behaviors~\cite{Hoogterp,Turk2002}. There are, however, other movements that do not augment a speech even though they appear frequently. Speakers show these movements habitually, without being aware of the body language during articulation---but they may nonetheless distract an audience during a speech ~\cite{ToastmastersInternational2011}. Toastmasters International dubbed these types of movements ``\emph{mannerisms}''~\cite{ToastmastersInternational2011}. Mannerisms include self-touching, scratching, gripping, leaning, finger-tapping, rocking, swaying, pacing,  fiddling with objects, adjusting hair and clothing, and more. Experts recommend avoiding mannerisms while delivering a speech~\cite{Hoogterp,Turk2002,ToastmastersInternational2011}.

Online public speaking tutoring systems (for example, ROCSpeak~\cite{Fung2015}) can help large numbers of people improve their verbal and non-verbal skills. Such systems make communication training affordable and accessible. An important building-block of these automated and scalable systems is the computational sensing and analysis of human behavior. We present a framework to computationally detect mannerisms from a multi-modal (audio-visual and ``MoCap,'' or motion capture, signals) record of public speaking. Systems capable of detecting mannerisms can make speakers aware of their body language and generate useful feedback. Analyzing the characteristics of mannerisms and where they appear will help further our understanding of fundamental aspects of human behavior.

Our current work stems from a previous attempt, called ``AutoManner''~\cite{Tanveer2015,Tanveer2016}, to detect repetitive body movements using an unsupervised algorithm. AutoManner only detects the repetitive movements and presents those to the users for their own judgments.  In our current work, we focus on classifying a repetitive movement as either a meaningful gesture or a mannerism.  This presents several challenges. For classification of body movements, it is crucial to understand the context of a speech. When a person gestures to compare two things, as in Figure~\ref{fig:comparison_gest}, it is meaningful. But that same gesture may be a mannerism in a different context. This is difficult to detect because we do not have an exhaustive list that links body movements with corresponding verbal contexts, nor can we easily or accurately infer a verbal context from a speaker's utterances. We address these challenges with the observation that mannerisms accompany hesitation. Relevant literature reveals that mannerisms and hesitations stem from the same source. As a result, correlation is likely. (See more discussion in Section~\ref{sec:prodsp}). We use this correlation to detect mannerisms by training the algorithm to take cues from speech hesitations \emph{and} body language.

\begin{figure}
  \centering
  \includegraphics[width=0.7\linewidth]{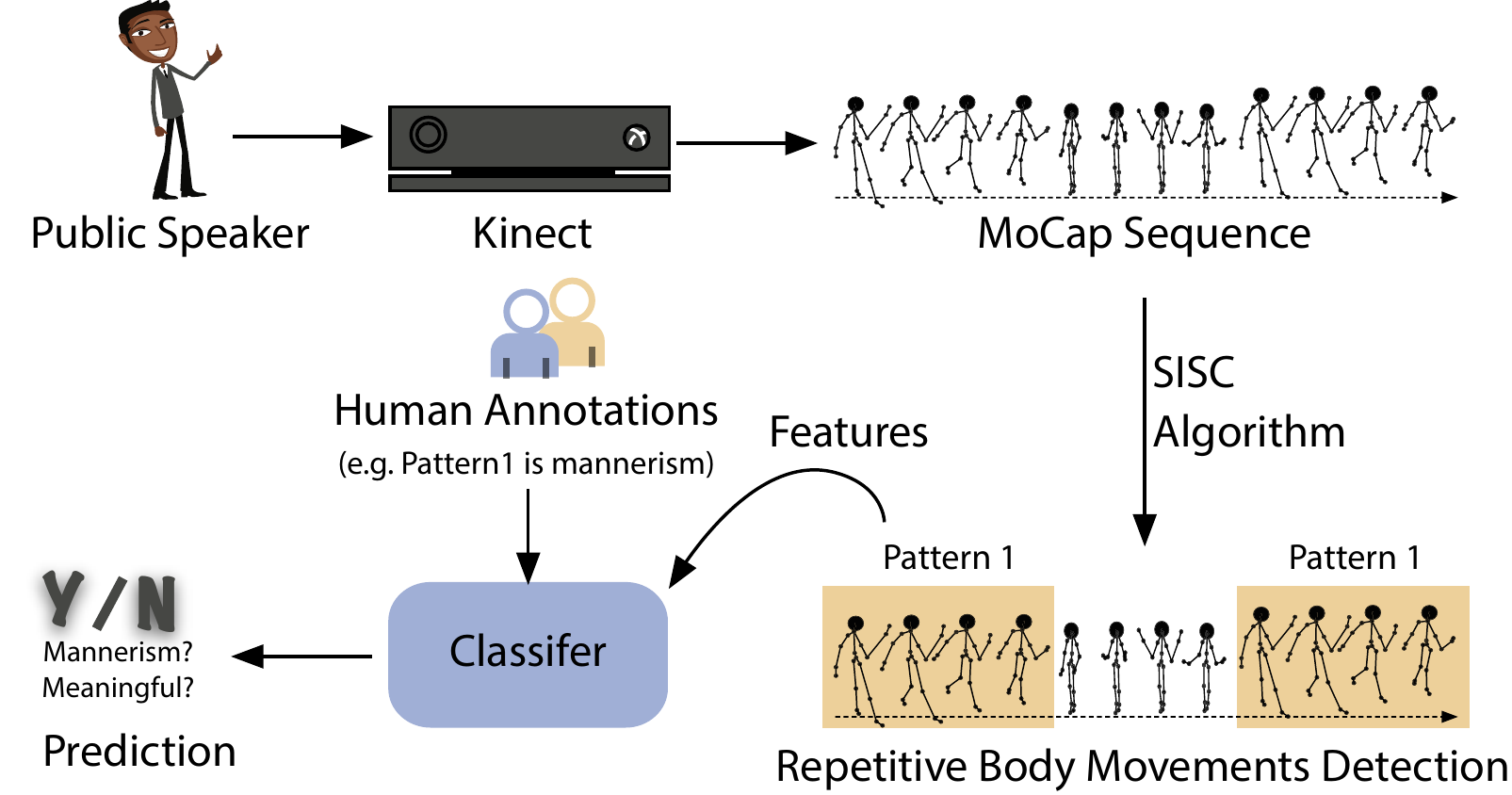}
  \caption{Detection of Mannerism in Public Speaking}
  \label{fig:main_diag}
\end{figure}

Our detection framework is illustrated in Figure~\ref{fig:main_diag}. We extracted MoCap sequences from 84 public speeches delivered by 28 speakers. A MoCap sequence records all the body movements of the speakers in three-dimensional coordinates. We use a \emph{shift-invariant sparse coding} (SISC) algorithm~\cite{Tanveer2015,Tanveer2016} to identify common body movement \emph{patterns}. A ``pattern'' is a short (approximately two-second) segment of a MoCap sequence that appears frequently within a speech. In this research, we are interested only in the frequent body movements, because, an uncommon movement is likely to be random. If a movement appears several times, it could be either meaningful or a mannerism, depending on the context.

Three online workers assess each pattern with the accompanying verbal context and decide whether the pattern is a mannerism or a meaningful gesture. We collected similar annotations from the public speakers themselves immediately after concluding their speeches. We extracted five different types of features and use several linear classifiers and regression methods to detect mannerisms. Our results show that it is possible to predict mannerisms with a much higher degree of accuracy than random chance. In addition, we analyzed the relative weight distributions in each classifier. Early analysis indicates that public speakers tend to focus more on \emph{what} they are saying, while audiences focus more on \emph{how} speakers say it. We released the dataset and code of our analysis.\footnote{The code of our analysis is available in https://github.com/ROC-HCI/AutoMannerPlus}

In sum, this paper details the following contributions:
\begin{itemize}
\item We propose a system to detect mannerisms using their concurrence with speech hesitations.
\item We design the system to predict mannerisms from both the speaker's perspective and the audience's perspective.
\item We quantify the differences between online workers' annotations and participants' self-annotations by evaluating the weight distributions of the trained classifiers. 
\item Our results indicate that speakers tend to focus more on \emph{what} they say while audiences focus more on \emph{how} speakers say it. 
\end{itemize}

\section{Related Literature}\label{sec:rellit}
To better understand mannerisms, we surveyed the relevant theory. In this section, we discuss the background literature on mannerisms, their relationship to speech hesitations, and other work similar to our research.
\subsection{Mannerisms and their Characteristics}
Body language is effective when it complements the verbal content it accompanies~\cite{Hoogterp,Turk2002}. When gestures are inconsistent with verbal content, they are referred to as ``mannerisms''~\cite{ToastmastersInternational2011}. Traditionally, gestures in speech communications are classified into four different categories: \emph{iconic}, \emph{metaphoric}, \emph{deictic}, and \emph{beat} gestures~\cite{McNeill2008}. Iconic gestures are hand movements illustrating object attributes or actions (demonstrating ``big'' by holding hands apart, for example). Metaphoric gestures illustrate abstract concepts more concretely (forming hands into a heart shape to represent love and affection). Deictic (or pointing) gestures are typically used to show the location of an object. Beat gestures reflect the rhythms of speech and are usually used to direct listeners' attention to important information~\cite{Biau2013}. Mannerisms, however, don't fit into these categories, as they are not meaningful to the content of the speech. They do not convey semantic information. They are expressed inadvertently to cope with the cognitive demands of public speaking~\cite{Grand1977,Harrigan1985}.

Mannerisms are distracting to audiences~\cite{Hoogterp,Turk2002}. Dick et al.~\cite{Dick2009} described an important phenomenon to partially explain why this happens. The group used functional magnetic resonance imaging (fMRI) to examine the influence of gestures on neural activity in brain regions associated with processing semantic information. Their experiment shows that the human brain is hardwired to look for semantic information in hand movements that accompany speech. In other words, listeners exhibit significant amounts of neural activity in relating body movements to verbal content. Mannerisms cognitively overburden audiences without conveying semantic information, amounting to nothing more than a distraction.

\subsection{Production of Speech Hesitations and Mannerisms}\label{sec:prodsp}
In our manual analysis, we noticed that mannerisms typically accompany speech hesitations, such as filler words and long pauses. In order to investigate the cause of this concurrence, we studied the existing literature.

Evidence suggests that speech hesitations occur when speakers are uncertain about what to say or when they need to decide from many choices~\cite{Corley2008}. For instance, Sharon Oviatt~\cite{Oviatt1995} observed that disfluencies more often occur before \emph{longer} utterances. The author showed that a simple linear regression over utterance length can account for 77 percent of variations in speech disfluency. Merlo et al.~\cite{Merlo2004} showed that disfluencies occur more when talking about unfamiliar topics. Beattie et al.~\cite{Beattie1979} observed that utterances with words described as "contextually improbable" are more likely to be disfluent. The group also suggested that speech disfluencies arise from an element of choice in selecting an appropriate word with low contextual probability. These studies suggest that disfluencies are likely to occur when speakers are burdened with thinking, planning, or choosing.

Stanley Grand~\cite{Grand1977} analyzed the characteristics and causes of a particular type of mannerism: \emph{self-touching}. He concluded that these hand movements are a form of feedback that helps speakers reduce cognitive overload by narrowing their attention. This helps speakers articulate simple chunks of information. Jinni Harrigan~\cite{Harrigan1985} performed a more recent study during medical interviews with 28 physicians and their patients. This study evinced that self-touching is related to information processing and production.

These experiments show that hesitation is a phenomenon observed in speaking situations with a high cognitive load. There is additional, more direct evidence that certain types of mannerisms are used to cope with high cognitive loads. As they arise in similar situations, it follows that both mannerisms and hesitations might appear together in a speech. We use this intuition to predict mannerisms by designing features related to speech hesitations---disfluency and prosody, for example.

\subsection{Similar Work}
Much research has been conducted to build systems to help with various aspects of public speaking. Damian et al. proposed a system named ``Logue''~\cite{Damian2015} to increase public speakers' awareness of their nonverbal behavior. It uses sensors to analyze speakers' speech rates, energy, and openness. Results are communicated to speakers using a head-mounted display. Roghayeh Barmaki~\cite{Barmaki2016} proposed an online gesture recognition application that would provide feedback through different channels, including visual and haptic channels. Bubel et al. created ``AwareMe''~\cite{Bubel2016} to offer feedback on pitch, use of filler words, and words per minute. AwareMe uses a detachable wristband to communicate with speakers while practicing. A Google Glass application named ``Rhema''~\cite{Tanveer2015a} provides real-time feedback to public speakers about their prosody (speech rate and volume). It was designed to reduce speaker distraction by providing succinct visual feedback. Similar efforts have been made by Luyten et al.~\cite{Luyten2016}, who explored the possibility of designing feedback systems by putting displays close to users' peripheral fields of vision. Such systems communicate real-time information to users more efficiently. ``ROCSpeak''~\cite{Fung2015} is an open and online semiautomatic system developed at the University of Rochester to provide feedback about a public speaker's performance. The system provides automated feedback on a speaker's loudness, intonation pattern, facial expressions, word usage, and more. It also collects and provides subjective evaluations from Amazon Mechanical Turk annotators.

There are several projects on detecting competence in public speaking. Chen et al.~\cite{Chen2014} proposed a multimodal sensing platform for scoring presentation skills. They used syntactic, speech, and visual features (e.g. hand movements, head orientations, etc.) with supervised regression techniques (support vector regression and random forest) to predict a continuous score for public speaking performance. They claimed a correlation coefficient of 0.38 to 0.48 with the manually-annotated ground truth. 

Only a few projects, however, focus on body language in public speaking. Nguyen et al.~\cite{Nguyen2012} implemented an online system to provide feedback on a speaker's body language. The feedback is given on a five-point scale. The authors recorded physical movements using a Kinect, then used a nearest-neighbor classifier to compare the recorded movements with a set of predefined templates of body movements. The templates contain ground truth information about the possible feedback. As it uses nearest-neighbor classification, this method cannot provide appropriate feedback for new (or unusual) body movements. Additionally, it cannot assess the contextual relationship of gestures to verbal content. 

In our previous work, we designed an interactive system, ``AutoManner''~\cite{Tanveer2016}, to make public speakers more aware of mannerisms. The system extracts  speakers' repetitive body movements in the form of patterns and shows the patterns to the speakers. The results reveal that question-and-answer--based interaction can make speakers aware of their mannerisms. The system could not, however, automatically detect which of the repetitive patterns were mannerisms. Without a detection mechanism, the users needed to review many repetitive patterns in order to identify their mannerisms.
\begin{figure}
\centering
\includegraphics[width=\linewidth]{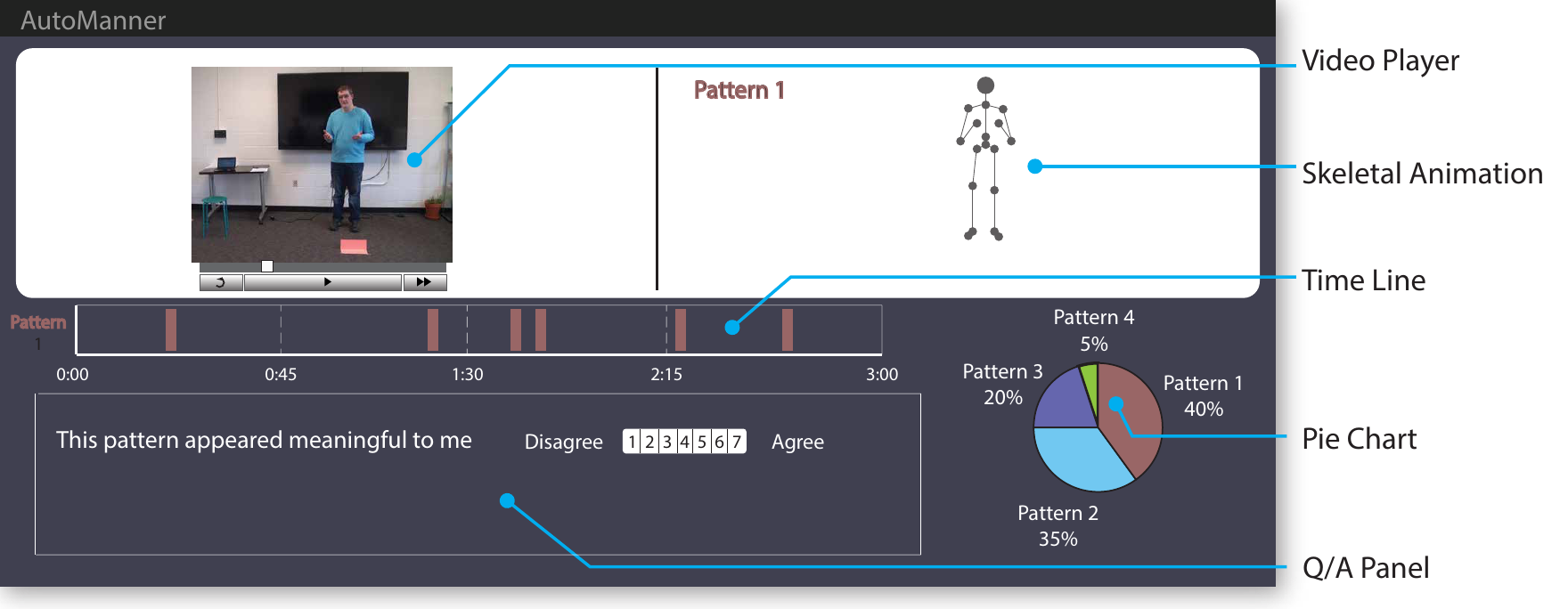}
\caption{Screenshot of the ground truth annotation interface}
\label{fig:screenshot}
\end{figure}

\section{Dataset}
In order to classify mannerisms, we use the AutoManner dataset, which we collected while evaluating the AutoManner system.\footnote{This dataset is available upon request in http://www.cs.rochester.edu/hci/currentprojects.php?proj=automanner.} The dataset consists of 84 public speeches from 28 speakers. The dataset is gender-balanced, with 14 female and 13 male participants. All speakers were undergraduate or graduate students of the University of Rochester, as well as native speakers of English. Each speaker spoke on three different topics for approximately three minutes each. The speakers chose their own topics with the general guideline that the topics should be easily understood by a general audience. The most common topics involved a favorite book, movie, computer game, hobby, superhero, celebrity or role model, passion, etc. Every speaker's topic was decided two days in advance to allow participants time to prepare. 

During the speech, the speaker's full body movements were recorded using a Kinect depth sensor. Kinect uses an infrared projector and camera arrangement to analyze the depth of a scene from its sensors. Kinect is often used as a video game component, associated most closely with Microsoft's Xbox 360 gaming console. We used the Microsoft Kinect SDK~\cite{shotton2013} to extract the three-dimensional coordinates of 20 joint locations on the subject's body---a MoCap sequence. We used the SISC algorithm ~\cite{Tanveer2015a} to extract common body movement patterns from the speeches. This algorithm is detailed in section~\ref{sec:algo}. We extracted 314 unique body movement patterns. 

The public speeches were recorded with a high-definition video camera. We manually transcribed the recordings, including all filler words (uh, um, ah, etc.).

\begin{figure}
\centering
\includegraphics[width=\linewidth]{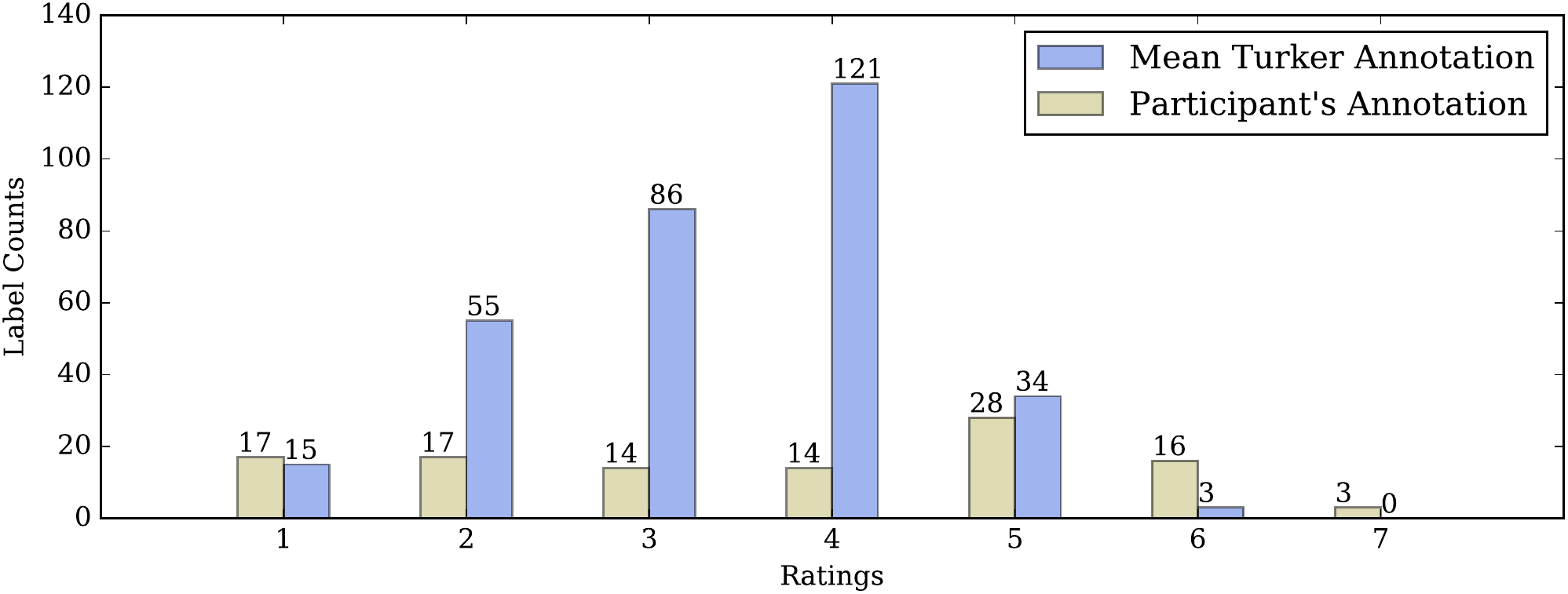}
\caption{Distribution of annotations}
\label{fig:label-dist}
\end{figure}
\subsection{Annotations}
After finishing the speech, each participant watched their own speech as part of the analysis phase. They used an interface similar to the one shown in Figure~\ref{fig:screenshot} to offer subjective annotations. The interface contains several components: a skeletal animation, a timeline, a video player, a pie chart, and a question-and-answer box. The pie chart shows how many patterns were extracted from the videos and their relative proportions. The skeletal animation shows the body movement patterns extracted by the SISC algorithm. The timeline highlights where a specific pattern appears during the speech. Participants can click on any of the highlighted regions on the timeline to play a video clip of the surrounding seconds of speech. This gives participants an opportunity to relate the content of the speech to the body movement patterns they are analyzing. Participants rated how meaningful their own body movement patterns were in the context of the speech. We collected ratings on a seven-point Likert scale, with seven indicating ``very meaningful.'' Before collecting the annotations, users were introduced to the interface and the analysis process with a pre-recorded demonstration video. 

The AutoManner study protocol required that only one video (of the three videos per subject) be annotated by the subjects themselves. That is not enough to classify mannerisms. In order to collect more ground truth data, we recruited online workers from Amazon's Mechanical Turk. The workers, known as ``Turkers," annotated the 315 patterns extracted from the database of 84 videos. For redundancy and more accurate measurement, we recruited three Turkers per video and averaged their annotations. The average ratings were quantized to fall within seven bins, ranging from one to seven. We used the interface shown in Figure~\ref{fig:screenshot} to gather annotations from the online workers. For quality assurance purposes, we selected only Turkers located in the United States who had completed at least 500 jobs and who had a 95 percent acceptance rate.

\subsection{Distribution of Annotations}
Figure~\ref{fig:label-dist} shows the distribution of the annotated labels over a range from one to seven. The distributions are shown for the average of the Turkers' annotations and the participant's annotations. The Turkers' average annotations are normally distributed due to the central limit theorem. Note that the total number of participant annotations is much smaller than the total number of Turker annotations because participants annotated only one video of the three.

\section{Research Questions}
Our research is focused to answer the following questions:
\begin{enumerate}
\item Is it possible to automatically detect mannerisms? If so, to what extent?
\item Is there any difference between self judgment and audience judgment in the mannerism annotations? If so, what is it?
\end{enumerate}
It is possible to manually observe the videos in the dataset to formulate a hypothetical answer to these questions. Manual analysis, however, is costly in terms of time, effort, and money. It is also difficult to reproduce. We perform a computational analysis of mannerisms. With this technique, it will eventually be possible to apply the same analysis to a new dataset, validating our results in new contexts. Computational analysis is also integral to our original objective of implementing an open, online platform for verbal and nonverbal skills training in public speaking.

\section{Technical Details}
In this section, we discuss various components of the computational framework, as outlined in Figure~\ref{fig:main_diag}.

\subsection{Extracting the Repetitive Patterns}\label{sec:algo}
\begin{figure}
\centering
\includegraphics[width=0.85\linewidth]{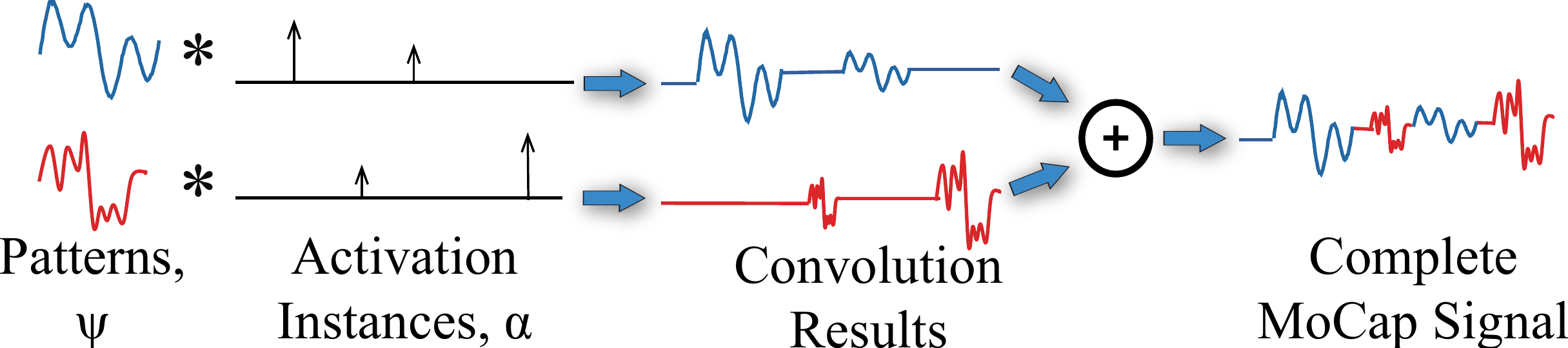}
\caption{Mathematical Model of the MoCap Sequence}
\label{fig:mocapmodel}
\end{figure}

We recorded the MoCap sequence of the participants' body movements using a Kinect sensor. We extracted the body movement patterns from the MoCap sequence using the SISC algorithm ~\cite{Tanveer2015,Tanveer2016}, an unsupervised algorithm for extracting frequently-occurring contiguous segments from a MoCap sequence.\footnote{SISC code is available in https://github.com/ROC-HCI/AutoManner} The algorithm does not require a list of templates for extracting the patterns. It works by solving the optimization problem shown in Equation~\eqref{eq:OptProb1}:
\begin{equation}
\label{eq:OptProb1}
    \begin{split}
        \hat{\psi}[m], \hat{\alpha}[n] = &  \arg\min_{\psi,\alpha}  \frac{1}{2}\|f[n]  - f_{\text{model}}[n]\|^2 +    \lambda\|\alpha\|_1\\
        \text{s.t. }\quad &\|{\psi}\|_F^2 \le 1 \quad \text{and,} \quad \forall_n\alpha[n] \geq 0.
    \end{split}
\end{equation}
In this equation,  $f[n]$ is a digital signal representing the MoCap signal captured from the Kinect sensor. Each sample of the signal is a 60-dimensional vector containing three-dimensional coordinates of 20 body joints. $f_\text{model}[n]$ represents a mathematical model of the MoCap signal, as shown in Equation~\eqref{eq:mocap_model}:
\begin{equation}
    \label{eq:mocap_model}
    \begin{split}
         f_\text{model}[n] = \sum_{d=0}^{D-1} \alpha_{d}[n] \ast \psi_d[m]
    \end{split}
\end{equation}
Here, $\psi_d$ represents one among a total of $D$ possible patterns. $\alpha_d$ represents the corresponding locations of $\psi_d$ within $f_\text{model}[n]$. $\alpha_d$ can be thought of as a sparse train of impulse functions. The asterisk represents the convolution operation. This model allows time-shifted replication of the patterns at the impulse locations, as shown in Figure~\ref{fig:mocapmodel}.

Minimizing the first term in the objective function---Equation~\eqref{eq:OptProb1}---ensures that the model parameters are adjusted in such a way that the MoCap model matches, as closely as possible, the actual signal captured by the Kinect. The $\ell_1$ norm of $\alpha$ (the second term in the objective function), however, makes the impulse train as sparse as possible. The constraints to the optimization ensure that $\alpha$ is non-negative and that the value of $\psi$ does not increase arbitrarily. Note that, although the overall optimization problem is non-convex, Equation~\eqref{eq:OptProb1} is convex when one of the two parameters, $\psi$ or $\alpha$, is considered constant. This allows us to solve this optimization problem with an alternating gradient descent approach, as described in Algorithm~\ref{algo:update}.
\begin{algorithm}[t]
\SetAlgoNoLine
 \KwIn{$\mathbf{f}[n]$, $M$, $D$ and $\lambda$}
 \KwOut{$\mathbf{\psi}$, $\mathbf{\alpha}$}
  \textbf{Initialize}\;
  $\mathbf{\alpha}\leftarrow 0$, $\mathbf{\psi} \leftarrow $ random\;
 \While{notConverge}{
      Update $\mathbf{\psi}$ using Gradient Descent. i.e. $\mathbf{\psi}^\text{new}\leftarrow \mathbf{\psi}^\text{old}-\gamma \mathop{\nabla}_{\psi}P$, where, $(P=\frac{1}{2}\|f[n]  - f_{\text{model}}[n]\|^2)$\;
      Project $\mathbf{\psi}$ into the feasible set, $\{\mathbf{\psi}:\|\mathbf{\psi}\|_\text{F}^2\le 1\}$\;
      Update $\mathbf{\alpha}$ using Gradient Descent. i.e. $\mathbf{\alpha}^\text{new}\leftarrow \mathbf{\alpha}^\text{old}-\gamma \mathop{\nabla}_{\alpha}P$\;
      Shrink $\alpha$ to enforce sparsity\;
      Project $\alpha$ to feasible set, $\{\alpha:\forall_n \alpha[n]\ge0\}$\;
 }
 \caption{Extracting the repetitive patterns}
 \label{algo:update}
\end{algorithm}
At every iteration, we alternatively hold $\alpha$ constant, updating $\psi$, and hold $\psi$ constant, updating $\alpha$. In addition, we \emph{shrink} the alpha values toward zero to enforce sparsity. Equation~\ref{eq:shrinkage} represents the shrinkage operation.
\begin{equation}\label{eq:shrinkage}
     \alpha[n] \leftarrow \text{sgn}(\alpha[n]) \text{max}(0,|\alpha[n]| - \gamma\lambda) \quad \forall_{0 \leq n < N}
\end{equation}
Here, $\gamma$ represents the learning rate of gradient descent. $\lambda$ represents the Lagrangian operator shown in Equation~\eqref{eq:OptProb1}. In our code, we heuristically set the length of the patterns, $\psi$, at two seconds. We set the maximum number of allowed patterns per signal, $D$, at five. 

\subsection{Feature Extraction}
We extracted five different categories of features to classify mannerisms. We extracted all features from the time frame in which a specific pattern appeared. We list these features in Table~\ref{tab:feats}. Below, we describe each feature and our rationale for including them.

\subsubsection{Disfluency Features} We noted earlier that mannerisms tend to appear alongside speech hesitations. Thus, to detect mannerisms, we designed features to capture hesitations. These features include the average number of times a speaker talks, uses filler words, remains silent, etc.---all while showing a particular pattern of body movements. The features adequately represent hesitation because speakers tend to use filler words or become quiet when hesitating. So, the average time spent speaking filler words or not speaking at all (silence) tends to increase relative to the mean time it takes to speak actual words during periods of hesitation. These features also capture the speaking speed, a good indicator of disfluency.

We compute these features by aligning the transcripts of speeches with the corresponding audio clips using the Penn Phonetics Lab forced aligner~\cite{Yuan2008}. The transcript contains all filler words. After alignment, we compute the average time the speaker takes to say a word, remain silent, or utter filler words while exhibiting a body movement pattern. We also calculate the counts of unique words, filler words, pauses, and their relative proportions. These features capture many important characteristics, including speaking rate, repetitions, etc.

\subsubsection{Prosody Features} Prosody features contain the intonation patterns, loudness, and formants of the vocal sound. It captures the speaking style of the speaker. Besides, it can be a good indication of the speaker's affective states---boredom, excitement, or confusion, for instance. We include prosody in our computational framework because confusion is a dominant cause of speech hesitation. Prosody has been found useful in much of the relevant research: intent modeling~\cite{Soman2009}, job interview performance prediction~\cite{Naim2015}, and more. We use PRAAT~\cite{Boersma2002} to extract the loudness, pitch, and the first three formants from the speech signal. Then, we compute summary statistics (mean, standard deviation, etc.) for the prosody signals from the two-second segments of each pattern.

\subsubsection{Body Movement Features} 
We extracted features related to pace and movement, as mannerisms are related to body language. We extracted the features from the MoCap sequences captured using the Kinect. We calculate the position, velocity, and acceleration of the speakers' elbows, wrists, knees, and ankles with respect to a reference point on their body. We calculate the mean and standard deviation of the velocity and acceleration. We also calculate the mean positions of the joint locations. These features capture a number of movements in each pattern.

\subsubsection{Facial Features} The face is perhaps the most important channel for non-verbal communication. Facial expressions encode many different affective states, including confusion and stress. As mannerisms stem from high cognitive load situations, facial expressions could be relevant for detecting mannerisms. We extract low-level facial movements to capture any possible information relevant to mannerisms. Facial features are useful in other research, as well~\cite{Rahman2012,Naim2015a}.
\begin{table}
\centering
\caption{List and counts of the extracted features}
\label{tab:feats}
\begin{tabular}{|c|c|c|}
    \hline
    Category & Feature Names & \makecell{Count} \\
    \hline
    Disfluency & \makecell[l]{Average time for uttering a word, filler word, and silences;\\ Count of words, filler words, and pauses; Relative proportions \\of words, filler words and pauses} & 9\\
    \hline
    Prosody & \makecell[l]{Avg., Min, Max, Range and Standard deviation of loudness, pitch;\\ $1^\text{st}$, $2^\text{nd}$, and $3^\text{rd}$ formants; Ratios of voiced to unvoiced regions} & 26\\
    \hline
    \makecell{Body \\Movements} & \makecell[l]{Mean and std of position, velocity, and acceleration of the elbows,\\wrists, knees, and ankles} & 40\\
    \hline
    Face & \makecell[l]{Pitch, Yaw, and Roll of head; Normalized distances\\ between various points on the face} & 24\\
    \hline
    Lexical & \makecell[l]{Counts of words in 23 LIWC categories} & 23\\
    \hline
    \end{tabular}
\end{table}

\begin{figure}
\centering
\includegraphics[width=0.35\linewidth]{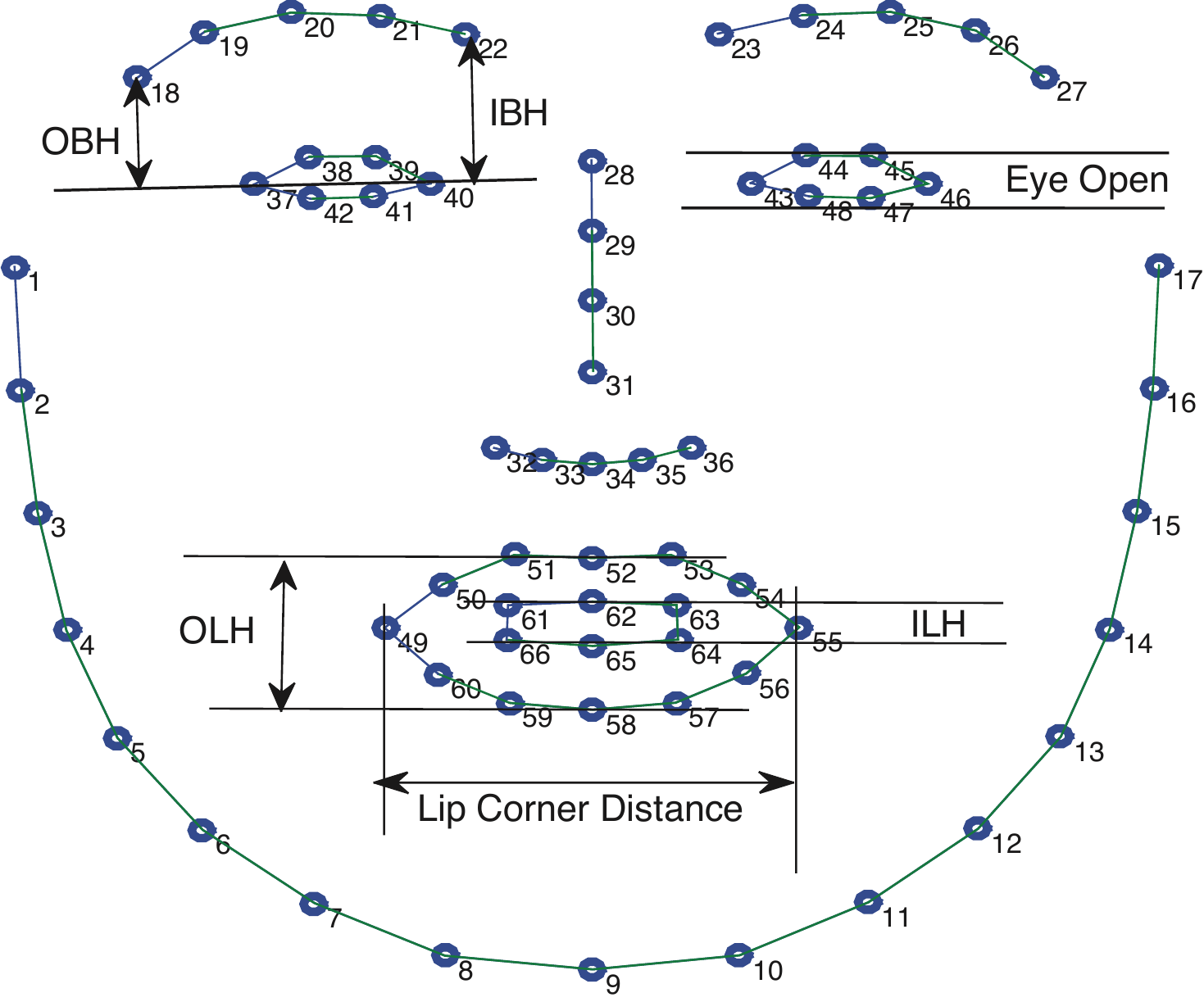}
\caption{Facial features extracted from the output of facial point tracker}
\label{fig:facial_feat}
\end{figure}

We use a facial point tracker, proposed by Saragih et al.~\cite{Saragih2011,Saragih2009}, to track 66 landmark points on the face. From these points, we calculate the pixel distances of OBH, IBH, OLH, ILH, eye-opening, and LipCDT, as illustrated in Figure~\ref{fig:facial_feat}. These distances are normalized by the pixel-wise distances between a subject's eyes to remove any scaling (zooming) effect. We compute the mean and standard deviation of these distances within the two-second length of each pattern. We also calculate the mean and standard deviation of the pitch, yaw, and roll movements of the head. The facial point tracker provides these estimated head movement measurements, which can reveal where a person is looking. In our manual observation, we found that looking away from the audience while gesturing repetitively is a good indicator of mannerisms. 

\subsubsection{Lexical Features} To capture information about the verbal content of the speech, we extract certain lexical features. To do this, we used a psycholinguistics tool, \emph{Linguistic Inquiry Word Count} (LIWC)~\cite{Pennebaker-liwc01}. LIWC describes 64 different categories of positive and negative emotions (happy, sad, angry, etc.), function word categories (articles, quantifiers, pronouns, etc.), content categories (anxiety, insight, etc.), and more. We use \emph{greedy backward elimination feature selection}~\cite{Caruana1994} to choose the 23 most relevant categories as lexical features.

\subsubsection{Feature Normalization} The features we select have different, dynamic ranges. To ensure that one group of features does not dominate the others, we apply z-score normalization over the features. That is, we subtract the mean of a particular feature from each feature value and divide by the standard deviation. This distributes the features with no resulting mean and unit variance.

\subsection{Classification Analysis}
To classify between mannerisms and meaningful gestures, we divide the patterns into two groups: all the patterns with a rating of four or higher are labeled meaningful, while patterns with a rating lower than four are labeled as mannerisms. This divides the dataset into two fairly equal groups. We use four different types of classification techniques: three linear classifiers (LASSO, LDA, and max-margin) to compare the relative proportions of the feature weights, as well as a nonlinear classifier (a neural network) to gauge improvement arising from a non-linearity in the classification. The feature weights yield valuable insights as to what makes a repetitive movements meaningful or, otherwise, mannerisms. 

\subsubsection{LASSO} The \emph{least absolute shrinkage and selection operator} (LASSO)~\cite{Tibshirani1996} is actually a linear regression technique with $\ell1$ norm regularization. The presence of the $\ell1$ norm enables it to perform variable selection and regression simultaneously. As we selected a large number of features from various channels of verbal and non-verbal communication, it is likely that some are correlated. LASSO should automatically choose the best combination of features and suppress unnecessary features. We made a minor change to LASSO to make it suitable for classification, as the original formulation of LASSO is designed for regression, by putting the logistic (Sigmoid) function over the linear predictor. Mathematically, we used the following formulation:
\begin{equation}
\label{eq:lasso}
    \min_{\beta}\left(\frac{1}{N}\|y-\sigma(\mathbf{X}\beta)\|_2^2\right)+\lambda\|\beta\|_1
\end{equation}
Here, $\sigma(t)=\frac{1}{1+e^{-t}}$ is the Sigmoid function, $\mathbf{X}$ is a matrix of all the features (columns) for all the datapoints (rows), $y$ is a vector containing $\{1,0\}$ (indicating the labels of the classes for each datapoint), and $\beta$ is a vector containing the coefficients for the linear regression line. We use Keras~\cite{Chollet2015} and Theano~\cite{2016arXiv160502688short} to solve the corresponding optimization problem.

\subsubsection{Max-Margin} The max-margin classifier, or \emph{support vector machine} (SVM)~\cite{Vapnik1964}, is a popular classification technique that maximizes the margin between two classes. It does not, however, perform automatic feature selection. We use the linear SVM implementation from the LibSVM library~\cite{Chang2011}. 

\subsubsection{LDA} Linear discriminant analysis (LDA)~\cite{Fisher1936} projects the data on to a maximally-separating hyperplane for classification. A maximally-separating hyperplane can emphasize the differences between meaningful and non-meaningful gestures, in terms of feature distributions. However, unlike LASSO, it does not perform feature selection. We use scikit-learn~\cite{Pedregosa2011} for both LDA classification and regression.

\subsubsection{Neural Network} Some features may depend on each other to represent mannerisms. Such interdependency is difficult to model with linear classifiers. We use a feed-forward neural network to capture a possible non-linear decision surface. In our network, we apply two hidden layers, each containing 16 nodes. The output layer contains only one node. We use a sigmoid or activation on the output node. We use Keras~\cite{Chollet2015} and Theano~\cite{2016arXiv160502688short} to implement the network.

\subsection{Regression Analysis}
We also performed regression analysis on the unmodified ratings of the classifier. Using regression analysis to predict the annotation ratings is unnatural because the Likert scale is not continuous. As a result, we do not expect high correlation with the predicted values. We performed the regression analysis anyway to compare different prediction techniques. Similar techniques as above were used. For the LASSO regression, we used the implementation from scikit-learn~\cite{Pedregosa2011}. For the  Support Vector Regression (SVR)~\cite{Smola2004} we used the libSVR library~\cite{Chang2011}. Scikit-Learn~\cite{Pedregosa2011} provides an interface for linear discriminant analysis--based regression. Finally, for neural network--based regression, we used the same architecture as the neural network classifier, except we use a \emph{Rectified Linear Unit} (ReLU) at the output layer, instead of sigmoid.

\section{Results}
In this section, we discuss the classification and regression performances of the predictors. We also discuss experiments we conducted to better understand our results.
\begin{figure}
\centering
\begin{minipage}{.45\textwidth}
  \begin{flushleft}
      \includegraphics[width=\linewidth]{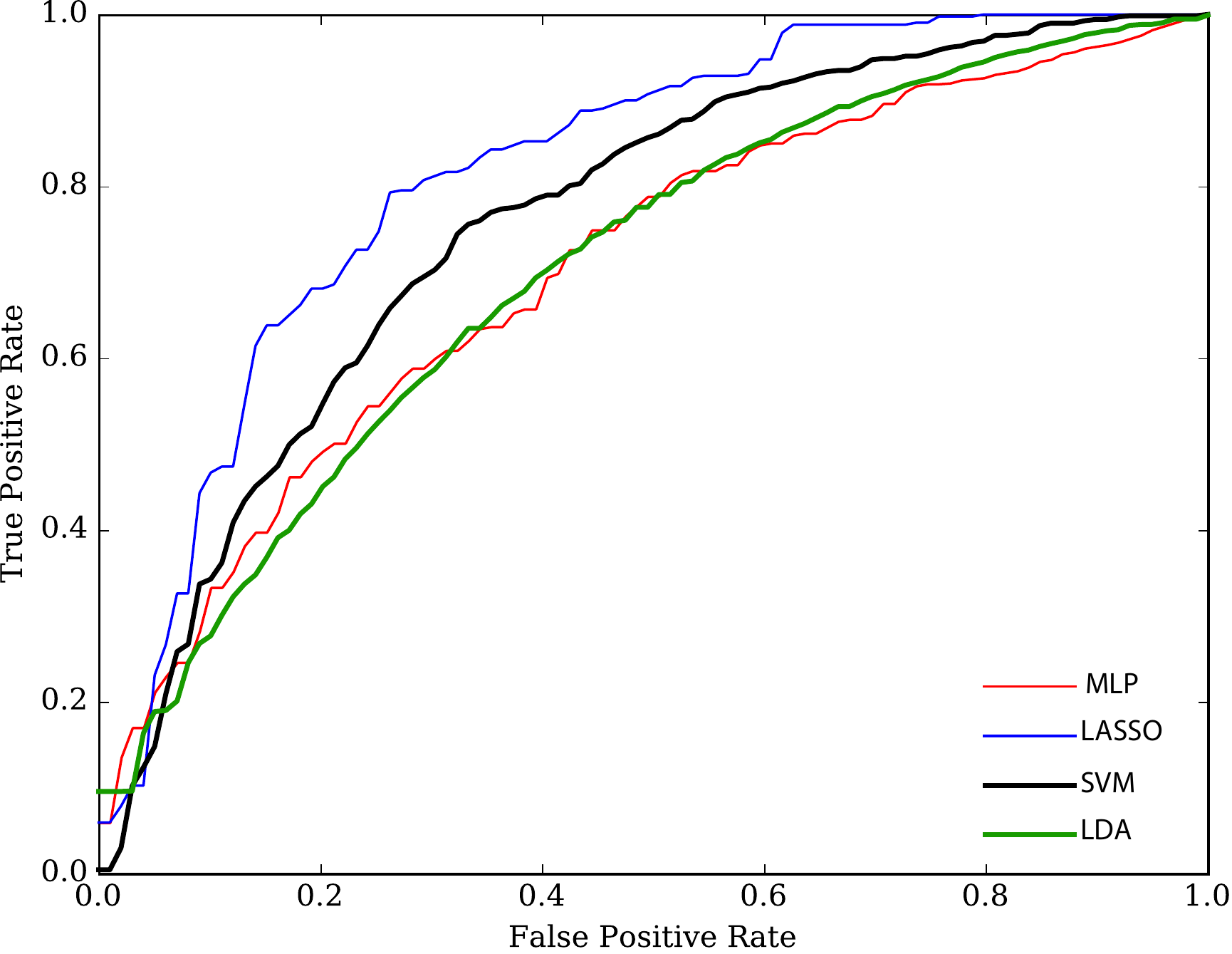}
      \caption{ROC Curves for classifiers trained over Mechanical Turk annotations}
      \label{fig:roc_mt}
  \end{flushleft}
\end{minipage}%
\hfill
\begin{minipage}{.45\textwidth}
  \begin{flushright}
      \includegraphics[width=\linewidth]{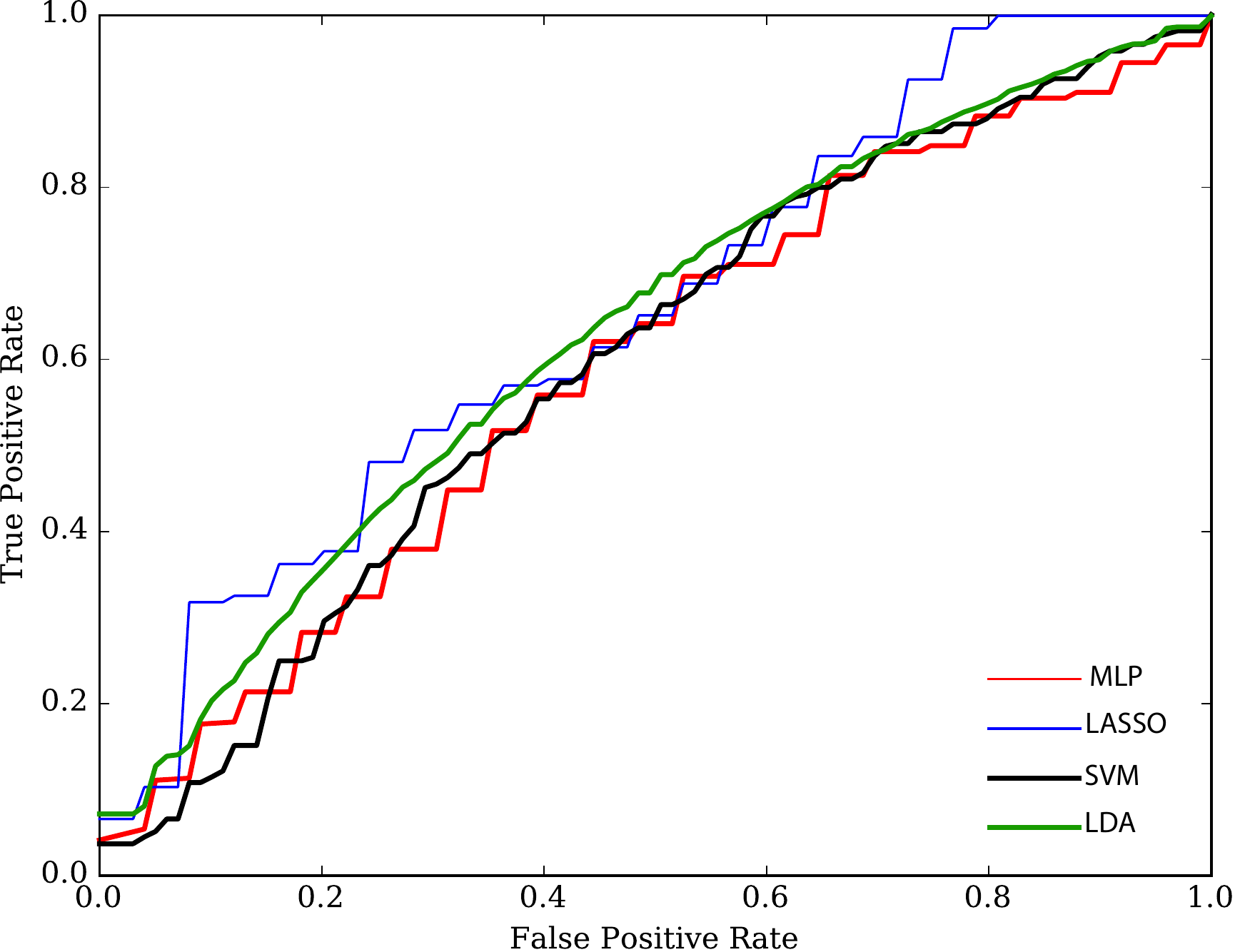}
      \caption{ROC Curves for classifiers trained over participant annotations}
      \label{fig:roc_part}
  \end{flushright}
\end{minipage}
\end{figure}

%\subsection{Inter-Rater Agreement}
%\begin{figure}
%\centering
%\includegraphics[width=0.4\linewidth]{Figures/agreements.pdf}
%\caption{Inter-Rater agreements among various annotators.}
%\label{fig:agreement}
%\end{figure}
%
%Figure~\ref{fig:agreement} shows the inter-rater agreement among the annotators. The agreement is calculated using Krippendorff's alpha~\cite{Krippendorff2004} and correlation coefficient. Krippendorff's alpha allows calculation of agreement among any number of raters, even with missing data and various types of measurements (binary, nominal, ordinal, interval, and more). In our experiment, we collect the annotation on a seven-point scale, which is an interval datum. Also, annotations are collected from more than two annotators. Therefore, Krippendorff's alpha is the ideal choice for measuring agreements in our experiment. The figure shows that there is moderate ($\alpha = 0.57$) agreement among the Mechanical Turk annotators. The individual Turkers, however, do not agree as much with the participants' annotations as they do among themselves. This provides the first indication that the Turkers' opinions were different than the participants' self-reflections.

\subsection{Classification Performances}
In Figure~\ref{fig:roc_mt}, we show the Receiver Operating Characteristics (ROC)~\cite{Hanley1982} of various classifiers that were trained using the Turkers' annotations. Figure~\ref{fig:roc_part} shows similar curves for the classifiers trained over participants' annotations. ROC curves are the plots of "True Positive Rates" vs. "False Positive Rates." To better compare, we calculated the Area Under the ROC (AUC)~\cite{Hanley1982}, shown in Table~\ref{classifier_auc}~(left). All ROC curves and areas below them were averaged across 30 measurements on randomly-assigned training and test subsets. The ``MTurk Annot.'' column represents the AUCs when we use all data with the Turkers' annotations. The classifier's performance on the data with the participants' self-annotated ratings is shown in the ``Self Annot.'' column. 

As there are only two classes in this task, a completely random classifier would have shown an AUC of 0.5. We observed a maximum AUC of 0.82. It is evident, therefore, that mannerisms can be classified with higher accuracy than a random classifier can achieve, at least for the case in which we use all the data from the Turkers' annotations as well as a LASSO classifier.

We became interested in other aspects of the result. Why are the participants' self-annotations so difficult to classify? (The maximum AUC is 0.65) The first explanation that comes to mind is the limited number of data available for training---the self-annotation dataset was approximately one-third the size of the Mechanical Turk dataset. Perhaps the limited number of data in the former dataset were not representative enough to successfully train a classifier. Therefore, for the sake of a fair comparison, we randomly sub-sampled the Mechanical Turk dataset to constitute a set one-third of the original size to train new classifiers. The AUCs of the classifiers on the sub-sampled data are shown in the ``Subsampled MTurk Annotation'' column.

\begin{table}
\centering
\caption{Results of various prediction approaches. (Left) Area Under the ROC Curve (AUC) of various classifiers; (Right) Correlation Coefficient for various regression methods.}
\label{classifier_auc}
\begin{tabular}{|c|c|c|c|}
    \hline
    Classif. &\makecell{ MTurk \\Annot.} & \makecell{Subsampled \\MTurk\\ Annot.} &
    \makecell{Self \\Annot.}\\
        \hline
        LASSO & \textbf{0.82} & 0.69 & \textbf{0.65}\\
        \hline
        \makecell{Max-\\Margin} & 0.78 & 0.69 & 0.60\\
        \hline
        LDA & 0.77 & \textbf{0.73} & 0.63\\
        \hline
        \makecell{Neural\\Network} & 0.71 & 0.60 & 0.59\\
        \hline
    \end{tabular}
\label{regression_corr}    
\begin{tabular}{|c|c|c|c|}
    \hline
    Regr. &\makecell{MTurk \\Annot.} & \makecell{Subsampled \\MTurk\\ Annot.} &         
    \makecell{Self \\Annot.}\\
        \hline
        LASSO & \textbf{0.63} & \textbf{0.55} & \textbf{0.37}\\ 
        \hline
        LDA & 0.59 & 0.38 & 0.31\\ 
        \hline
        \makecell{Neural\\Network} & 0.48 & 0.35 & 0.28\\ 
        \hline
        \makecell{Max-\\Margin} & 0.35 & 0.20 & 0.05\\ 
        \hline
    \end{tabular}
\end{table}
 It is evident from Table~\ref{classifier_auc}~(left) that for all the classifiers, even when the sub-sampled Turkers' dataset is used, the classifiers performed poorly for the self-annotation dataset. We performed statistical t-tests between the prediction results of self-annotations and sub-sampled Mechanical Turk annotations. The differences between these two groups are statistically significant: $p<<0.01$ in each case.  Therefore, we think the results provide a strong indication that the difference between self-annotations and Turkers' annotations is not an anomaly arising from the quantity of data. Participants' annotations seem to be qualitatively different than the Turkers' annotations. We discuss this effect more in Section~\ref{sec:disc}.

\subsection{Regression Performances}
To measure regression performance, we calculated the correlation coefficient between the regression output and the human annotations. Table~\ref{regression_corr}~(right) shows the performance of the regression methods in predicting mannerisms. The maximum correlation coefficient we obtained is 0.63. The regression results show a similar trend as in Table~\ref{classifier_auc}~(left): The Mechanical Turk annotations are more predictable than self-annotations. LASSO still performs best at the regression task. Max-margin's performance is particularly poor. Specifically, while predicting the self-annotations, max-margin's regression performance is nearly equivalent to a random predictor. We discuss this more in Section~\ref{sec:disc}.

\subsection{Distribution of weights}
The results indicate that the speakers' self-judgments are different than the audiences' (the Turkers') opinions. In order to understand the qualitative difference between these two, we look into the feature weights assigned by the linear predictors (classifiers and regressors). Analyzing the weights helps us identify how various features contribute to the prediction task. We calculate the weights of each feature-category (prosody, disfluency, etc.) using Equation~\eqref{eq:weight}:
\begin{equation}
W_c=\frac{1}{N_c}\sum_{f\in c}{|W(f)|}
\label{eq:weight}
\end{equation}
Here, $c$ is a feature category, $N_c$ is the number of features with nonzero weights in $c$, and $W(f)$ is the weight of the feature $f$. The names of the features are given in Table~\ref{tab:feats}.

\begin{figure*}
\centering
\includegraphics[width=\linewidth]{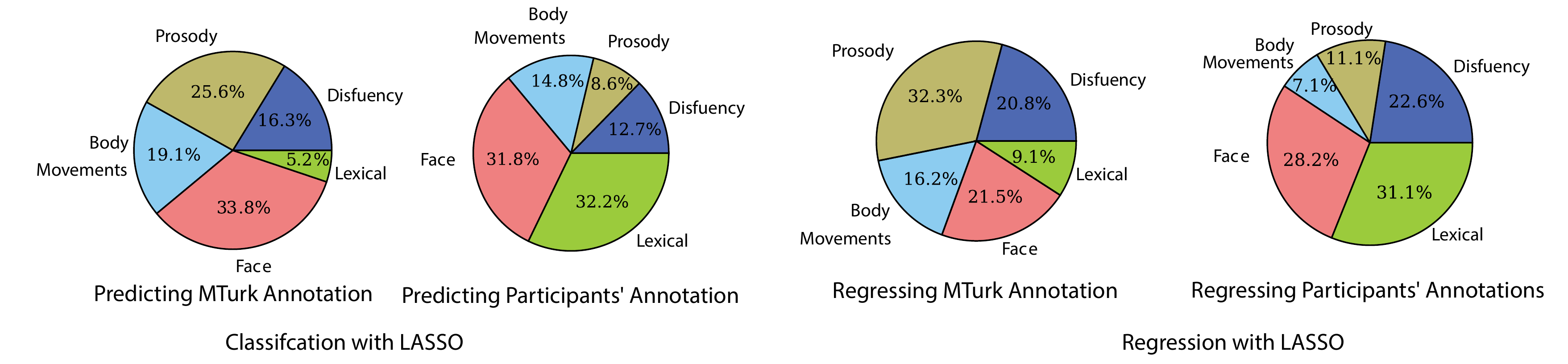}
\caption{Distribution of various types of feature weights in the trained LASSO classifier}
\label{fig:coef-dist}
\end{figure*}

In Figure~\ref{fig:coef-dist}, we show the relative proportions of these normalized weights in pie charts. Notice that the lexical features take only 5.2 percent of the total weights to classify the Turkers' annotations, but 32.2 percent to classify the participants' annotations. A similar trend is visible in the regression task. Lexical features contribute only 9.1 percent of the total weights for Turkers' annotations. The same features take 31.1 percent of the weights while regressing the speakers' self-annotations. In other words, lexical features are more predictive of the participants' annotations than the Turkers' annotations. Notice, however, that the prosody and body movement features reveal an opposing trend. They contribute more in predicting the Turkers' annotations, but less in predicting the speakers' annotations.

We observe the weights of the other two linear classifiers, max-margin and LDA, to ensure the results aren't coincidental. Table~\ref{tab:weights} shows the distributions of the feature weights in varied cases. ``MTF'' and ``MTS'' represent prediction of all Mechanical Turk annotations and their random sub-samples. ``Self'' represents prediction of the speakers' self-annotations. Note that, in most cases, lexical features take a higher percentage for self-annotations than the Turkers' annotations. This is consistent with earlier observations. The only exception is when max-margin is used for regression purposes. We discussed previously, of course, that max-margin was a poor regressor for predicting self annotations. Assigning comparatively lower weights on the lexical feature might be responsible for the poor performance.

This trend in feature-weight distribution suggests an interesting phenomenon: participants are perhaps more observant about the verbal content of their own speeches whereas online workers primarily notice the speakers' body language. This hypothesis seems plausible if we consider how much effort speakers exert to produce their speeches. 
\begin{table*}
\centering
\caption{Percentages of Weights in Various Classification and Regression Techniques}
\label{tab:weights}
\centering
\begin{tabular}{|*{8}{c|}}
\hline
\multirow{2}{*}{Method}     & \multirow{2}{*}{Task}           & \multirow{2}{*}{Annot.} & \multicolumn{5}{l|}{Percentages of Weights for various Feature Categories} \\ \cline{4-8} 
                            &                                 &                             & Disf.(\%)     & Prosody(\%)     & Body(\%)    & Face(\%)    & Lexical(\%)    \\ \hline
\multirow{6}{*}{\makecell{Max-\\Margin}} & \multirow{3}{*}{Classif.} & MTF                  & 10.4           & 26.3        & 17.7    & 26.4    & 19.3       \\
                            &                                 & MTS            & 11.9           & 39.9        & 16.1    & 15.5    & 16.6       \\  
                            &                                 & Self                        & 18.3           & 13.0        & 7.4     & 26.9    & 34.5       \\ \cline{2-8} 
                            & \multirow{3}{*}{Regr.}     & MTF                  & 20.7           & 19.1        & 16.8    & 21.1    & 22.3       \\ 
                            &                                 & MTS            & 27.8           & 16.8        & 13.9    & 18.6    & 22.8       \\
                            &                                 & Self                        & 21.4           & 18.8        & 16.8    & 22.0    & 20.9       \\ \hline
\multirow{6}{*}{LDA}        & \multirow{3}{*}{Classif.} & MTF                  & 19.0           & 20.5        & 17.5    & 22.6    & 20.3       \\
                            &                                 & MTS            & 20.1           & 21.9        & 14.6    & 21.6    & 21.7       \\
                            &                                 & Self                        & 20.9           & 17.0        & 14.1    & 23.2    & 24.8       \\ \cline{2-8} 
                            & \multirow{3}{*}{Regr.}     & MTF                  & 21.5           & 19.6        & 19.0    & 19.5    & 20.4       \\
                            &                                 & MTS            & 22.3           & 19.2        & 17.0    & 18.8    & 22.7       \\
                            &                                 & Self                        & 20.9           & 16.8        & 15.5    & 21.4    & 25.4       \\ \hline
\end{tabular}
\end{table*}

\section{Discussions}\label{sec:disc}
The results of our experiment support the argument that speakers' annotations and Turkers' annotations are different. \textbf{First}, the performances of the predictors reveal that the Turkers' annotations are much easier to predict than the self-annotations. They might, therefore, be qualitatively different. \textbf{Second}, the distribution of weights strongly indicates a difference in how the features contribute to predicting the two different kinds of annotations. The speakers emphasize verbal features more in determining if a gesture is a mannerism. The audience, on the other hand, primarily emphasizes non-verbal aspects (prosody and body movement, among others) in detecting mannerisms. This might be another reason speakers are typically unaware of their own body language.

The distribution of feature weights also provides a clue as to why the classification and regression techniques performed poorly in predicting self-annotations. The weights imply that the participants focus less on the non-verbal aspects of the speech, putting more emphasis on the verbal aspects. We did not, however, extract many verbal features in this experiment. The LIWC features capture only a statistical distribution of the words in a ``Bag of Words'' model. It does not capture any syntactic information. Many important mannerism cues might be captured in grammatical accuracy, sentence formulation, or stylistic aspects of the speech. Indeed, it is possible that the features we selected did not include the full spectrum of verbal qualities. It is, however, also possible to capture additional nonverbal features including eye contact, high-level interpretation of facial expressions (i.e. gauging surprise, happiness, concentration, thought, etc.), and characteristics of pauses (using pauses to build suspence, for example). Future work will involve adding more verbal and nonverbal features to further validate further our findings.

Although LASSO and max-margin performed comparably well in the classification task, LASSO far outperforms max-margin for regression. In most cases, LASSO performed better than the other techniques, too. Note that another prominent characteristic of the weight distribution is that facial and disfluency features are consistently good predictors of mannerisms. We deliberately designed the disfluency (speech hesitation) features to predict mannerisms well. We did not appreciate the power of facial features until this point. In retrospect, it seems natural that the facial features work well, as they are one of the major nonverbal cues used by humans.

%\section{Limitations}
We want to emphasize that although our results \emph{support} the hypothesis that ``speakers tend to focus more on what they say while the audience focuses more on how the speakers say it,'' they certainly do not prove it. Instead, our results should be considered interesting observations from one particular dataset which needs substantiation from other public speaking datasets.

\section{Conclusion}
In this work, we proposed a computational approach to automatically detect mannerisms during a speech. This method detects mannerisms with reasonable accuracy (AUC up to 0.82). The proposed system can also detect mannerisms from both the speakers' and the audiences' perspectives. This system can be useful in making public speakers aware of their body language. 

Deeper analysis of the prediction methods yielded interesting insights on human behavior. Our results indicate that the way a speaker evaluates their own speech is different than the way an audience might. Speakers tend to emphasize the verbal components of the speech while the audience focuses more on non-verbal aspects. This finding could be useful in designing a new type of feedback to rethink assessment technologies and public speaking. 

\section{Acknowledgement}
This work was supported in part by Grant W911NF-15-1-0542 with the US Defense Advanced Research Projects Agency (DARPA) and the Army Research Office (ARO). Special acknowledgment to Microsoft Azure for providing the computational platform. Thanks to Vivian Li for preparing the first figure.

\bibliographystyle{abbrv}
\bibliography{tanveer_arxiv}  

%%%%%%%%%%%%%%%%%%%%%%%%%%%%%% End of Body %%%%%%%%%%%%%%%%%%%%%%%%%%%%%%%%%

\end{document}